Chapter 1

# NODE CLASSIFICATION IN SOCIAL NETWORKS

Smriti Bhagat
*Rutgers University*
smbhagat@cs.rutgers.edu

Graham Cormode
*AT&T Labs–Research*
graham@research.att.com

S. Muthukrishnan
*Rutgers University*
muthu@cs.rutgers.edu

**Abstract** When dealing with large graphs, such as those that arise in the context of online social networks, a subset of nodes may be labeled. These labels can indicate demographic values, interest, beliefs or other characteristics of the nodes (users). A core problem is to use this information to extend the labeling so that all nodes are assigned a label (or labels).

In this chapter, we survey classification techniques that have been proposed for this problem. We consider two broad categories: methods based on iterative application of traditional classifiers using graph information as features, and methods which propagate the existing labels via random walks. We adopt a common perspective on these methods to highlight the similarities between different approaches within and across the two categories. We also describe some extensions and related directions to the central problem of node classification.

## 1. Introduction

The emergence of online social networks (OSNs) in the past decade has led to a vast increase in the volume of information about individuals, their activities, connections amongst individuals or groups, and their opinions and thoughts. A large part of this data can be modeled as *labels* associated with individuals, which are in turn represented as nodes within a graph or graph-like structure. These labels come in many forms: demographic labels, such as age, gender and location; labels which represent political or religious beliefs; labels that encode interests, hobbies, and affiliations; and many other possible characteristics capturing aspects of an individual's preferences or behavior. The labels typically appear on the user's profile within the network, or attached to other objects in the network (photos, videos etc.).

There are many new applications that can make use of these kinds of labels:

- Suggesting new connections or contacts to individuals, based on finding others with similar interests, demographics, or experiences.
- Recommendation systems to suggest objects (music, movies, activities) based on the interests of other individuals with overlapping characteristics.
- Question answering systems which direct questions to those with most relevant experience to a given question.
- Advertising systems which show advertisements to those individuals most likely to be interested and receptive to advertising on a particular topic.
- Sociological study of communities, such as the extent to which communities form around particular interests or affiliations.
- Epidemiological study of how ideas and "memes" spread through communities over time.

Of course, these are just a few examples of the many different ways social network data is of interest to businesses, researchers, and operators of social networks. They have in common the aspect that knowing labels for individuals is a key element of each application.

In an ideal world (as far as these applications are concerned), every user within a social network is associated with all and only the labels that are relevant to them. But in the real world, this is far from the case. While many users choose labels to apply to themselves, these labels can be misleading, inappropriate, outdated, or partial. This is for a variety of reasons: users may fail to update or add new labels as time progresses, letting their profile information grow "stale"; out of concerns for privacy, users may omit or distort information about themselves; users may simply forget or neglect to include information about their most important activities and interests; and some users simply delight in listing wantonly misleading information to amuse themselves and those who know them. Such distortions are prevalent, although not overwhelmingly so [18]. The consequence of this noise is to reduce the effectiveness of methods for the applications listed above. In particular, the most pressing problem is the absence of labels in certain categories (such as demographics or interests), which can make it impossible to provide useful suggestions or recommendations to that user.

**The Node Classification Problem.** This leads to the central problem of interest in this chapter: given a social network (or more generally, any network structure) with labels on some nodes, how to provide a high quality labeling for every node? We refer to this as the "node classification problem", with the understanding that the basic problem can be abstracted as providing a labeling for nodes in a graph structure. Variations on this problem might work over generalized graph structures, such as hypergraphs, graphs with weighted, labeled, or timestamped edges, multiple edges between nodes, and so on.

A first approach to this problem is to engage experts to provide labels on nodes, based on additional data about the corresponding individuals and their connections. Or individuals can be incentivized to provide accurate labels, via financial or other inducements. Indeed, historically this is exactly what sociologists have done when studying social groups of the order of tens to a hundred nodes, for example [35]. But this approach does not scale when confronted with networks containing hundreds of thousands to millions of individuals. While it may be feasible to rely on a moderate number of expert labeled nodes, or even a large fraction of "noisy" self-labeled nodes, this is very far from the goal of all nodes perfectly labeled.

Instead, we consider methods which use the information already encoded in the partially labeled graph to help us predict labels. This is based on the paradigm of machine learning and classification. In other words, we aim to train a classifier based on the examples of nodes that are labeled so we can apply it to the unlabeled nodes to predict labels for them (and also to nodes that have some labels to augment or replace their current labels). However, there are several aspects of this setting which make it somewhat different to the traditional model of classification, as will become apparent.

As is usual in machine learning, we first have to identify some "features" of nodes that can be used to guide the classification. The obvious features are properties of the node itself: information that may be known for all (or most) nodes, such as age, location, and some other existing nodes. But the presence of an explicit link structure makes the node classification problem different from traditional machine learning classification tasks, where objects being classified are considered independent. In contrast to the traditional setting, we can define additional features, based on adjacency or proximity in the graph. A first set of features are based on simple graph properties: the degree (number of neighbors) of the node; the neighborhood size reachable within two or three steps; the number of shortest paths that traverse through the node, and so on. But perhaps more interesting are features derived from properties of the nearby nodes: the *labels* of the neighbors form a canonical feature in this setting.

One may ask why the labels of neighboring nodes should be useful in predicting the label of a node. Certainly, if the edges between nodes were completely arbitrarily generated, there would not be much information to glean. But in social networks, links between nodes are far from arbitrary, and typically indicate some form of a relationship between the individuals that the nodes represent. In particular, a link can indicate some degree of similarity between the linked individuals: certainly not exact duplication, but sufficient to be a useful input into a learning algorithm.

Formally, the social sciences identify two important phenomena that can apply in online social networks:

- **homophily**, also known informally as "birds of a feather", is when a link between individuals (such as friendship or other social connection) is correlated with those individuals being similar in nature. For example, friends often tend to be similar in characteristics like age, social background, and education level.

- **co-citation regularity** is a related concept, which holds when similar individuals tend to refer or connect to the same things. For example, when two individuals have similar tastes in music, literature or fashion, co-citation regularity suggests that they may be similar in other ways or have other common interests.

If one can argue that either of these phenomena apply in a network of interest, then it suggests that information about nodes with short graph distance, or with similar attributes, may be useful in helping to classify a node of interest.

A secondary aspect of the graph setting is that the classification process can be iterative. That is, we may be faced with a node such that we initially have very little information about the node or its neighborhood. However, after an initial application of classification, we may know have a richer set of (putative) information about the neighborhood, giving more of a basis to label the node in question. In other words, the classification process can spread information to new places in the graph, and feed into the features that we use for classification. The classification continues to be iterated until it converges, or a fixed number of iterations have taken place. This iterative approach stands in contrast to the traditional model of classification, where the feature set is given at the start, and does not alter. The use of the graph edges to spread the labeling can be thought of as a special case of semi-supervised learning (where both labeled and unlabeled examples are used to build the classifier).

**Chapter Outline.** In this chapter we survey techniques that have been proposed to address the problem of node classification. We consider two broad classes of approaches.

In the first, we try to build on the vast knowledge of methods to solve the traditional classification problem. In other words, we define a method to generate vectors of features for each node (such as labels from neighboring nodes), and then apply a "local classifier" such as Naïve Bayes, decision trees and so on to to generate the labeling. As indicated above, this approach can be *iterative*: after a first round of classification, by training on the newly labeled examples.

Many other techniques have been suggested that more directly use the structure of the graph to aid in the labeling task. We take a unifying approach, and observe that it is possible to view many of these methods as performing *random walks* over the network to determine a labeling function. We thus refer to these as random walk based methods, and compare them in a common framework. This helps to identify similarities and highlight differences between these techniques. In particular, we see that all methods can be described via similar iterative matrix formulations, and that they are also closely related to iterative approaches with simple classifiers.

Based on this taxonomy of methods, we proceed as follows: in Section 2 we formalize the notions of graphs and labelings, and more precisely define the node classification problem over these structures. Given these definitions, we present methods for this problem in the graph domain. Then Section 3 describes the (iterative) local classifier method, while Section 4 explains how many methods can be viewed as random walks. We present additional details on applying the methods to large social networks in Section 5. Sections 6 and 7 discuss other approaches and related problems respectively, and we give concluding remarks in Section 8.

## 2. Problem Formulation

In this section, we discuss how online social networks (or more generally other networks) can be represented as (weighted) graphs. We then present the formal definition of the node classification problem.

### 2.1 Representing data as a graph

We consider data from social networks such as Facebook and LinkedIn, as well as other online networks for content access and sharing, such as Netflix, YouTube and Flickr. As is standard in this area, we choose to represent these networks as graphs of nodes connected by edges. In our setting, we consider graphs of the form $G(V, E, W)$ from this data, where $V$ is the set of $n$ nodes, $E$ is the set of edges and $W$ is the edge weight matrix. We also let $\mathcal{Y}$ be a set of $m$ labels that can be applied to nodes of the graph.

Given a network such as those above, there are typically many choices of how to perform this modeling, depending on which features are captured by the model and which are overlooked. The question of modeling network data is a topic worthy of study on its own. Likewise, the question of how this data is collected and prepared (via random sampling, crawling, or activity monitoring) is beyond the scope of this survey; rather, for this presentation we assume that the data is readily available in the desired graph model. For clarity, we provide some illustrative examples of how (social) network data may be captured by a variety of choices of graph models:

Example 1.1 *Consider Facebook as an example of a modern, complex social network. Users of Facebook have the option of entering a variety of personal and demographic information into their Facebook profile. In addition, two Facebook users may interact by (mutually) listing each other as friends, sending a message, posting on a wall, engaging in an IM chat, and so on. We can create a graph representation of the Facebook social network in the form* $G(V, E, W)$*, where*

- *Nodes* $V$*: The set of nodes* $V$ *represents users of Facebook.*
- *Edges* $E$*: An edge* $(i, j) \in E$ *between two nodes* $v_i, v_j$ *could represent a variety of possibilities: a relationship (friendship, sibling,* *partner), an interaction (wall post, private message, group message), or an activity (tagging in a photo, playing games). To make this example concrete, consider only edges which represent declared "friendships".*

- *Node Labels $\mathcal{Y}$: The set of labels at a node may include the user's demographics (age, location, gender, occupation), interests (hobbies, movies, books, music) etc. Various restrictions may apply to some labels: a user is allowed to declare only one age and gender, whereas they can be a fan of an almost unlimited number of bands.*

- *Edge Weights $W$: The weight $w_{ij}$ on an edge between nodes $v_i, v_j$ can be used to indicate the strength of the connection. In our example, it may be a function of interactions among users, e.g., the number of messages exchanged, number of common friends etc.; or it may simply be set to 1 throughout when the link is present.*

EXAMPLE 1.2 *As an example of a different kind of a network, consider the video sharing website, YouTube. Let graph $G(V, E, W)$ represent the YouTube user network, where*

- *Nodes $V$: A node $v_i \in V$ represents a user.*

- *Edges $E$: An edge $(i, j) \in E$ between two nodes $v_i, v_j$ could be an explicit link denoting subscription or friend relation; alternately, it could be a derived link where $v_i, v_j$ are connected if the corresponding users have co-viewed more than a certain number of videos.*

- *Node Labels $\mathcal{Y}$: The set of labels at a node may include the user's demographics (age, location, gender, occupation), interests (hobbies, movies, books, music), a list of recommended videos extracted from the site, and so on.*

- *Edge Weights $W$: The weight on an edge could indicate the strength of the similarity by recording the number of co-viewed videos.*

EXAMPLE 1.3 *Using the same YouTube data, one can derive a quite different graph $G(V, E, W)$, where*

- *Nodes $V$: A node $v \in V$ represents a video.*

- *Edges $E$: An edge $(i, j) \in E$ between two nodes $v_i, v_j$ may represent that the corresponding videos are present in certain number of playlists or have been co-viewed by a certain number of people.*

- *Node Labels $\mathcal{Y}$: The set of labels at a node may be the tags or categories assigned to the video, the number of times viewed, time of upload, owner, ratings etc.*
- *Edge Weights $W$: The weight on an edge may denote the number of users who co-viewed the videos.*

The graphs abstracted in each example vary not only in the semantics of nodes, edges and labels but also in graph properties such as directionality, symmetry and weight. For instance, the friend interactions in Facebook and YouTube are reciprocal, hence a graph $G$ where edges represent friendship relationship is *undirected*, i.e. it has a *symmetric* adjacency matrix. On the other hand, a graph where an edge represents a subscription in YouTube or a wall post in Facebook, is directed. Further, depending on the availability of interaction information, the graph may be *unweighted*, and all edges treated uniformly.

**Inducing a graph.** In some applications, the input may be a set of objects with no explicit link structure, for instance, a set of images from Flickr. We may choose to *induce* a graph structure for the objects, based on the principles of homophily or co-citation regularity: we should link entities which have similar characteristics (homophily) or which refer to the same objects (co-citation regularity).

Applying these principles, several choices of how to induce graphs have been advocated in the literature, including:

- Materializing the fully connected graph on $V$ where each edge $(i, j)$ is weighted by a suitable distance metric based on the similarity between $v_i$ and $v_j$. A special case is the exp-weighted graph where the weight on an edge $(i, j)$ is given by:

  $$w_{ij} = \exp\left(-\frac{\|v_i - v_j\|^2}{2\sigma^2}\right) \tag{1.1}$$

  Here, nodes in $V$ are interpreted as points in a geometric space (such as Euclidean space) and $\sigma^2$ represents a normalizing constant, which is the the variance of all points.

- Given a node $v_i$, let $NN(v_i)$ denote the set of $k$ nearest neighbors, i.e. the $k$ other nodes which are closest based on some measure of distance: this could be cosine or euclidean distance based on the features of the nodes. The $k$ Nearest Neighbor ($k$NN) graph then has an edge between a pair of nodes $v_i, v_j$ if $v_j \in NN(v_i)$. A $k$NN graph is by definition a directed graph with each node having outdegree $k$. An undirected, symmetric $k$NN graph can be defined where nodes $v_i, v_j$ are connected if $v_i \in NN(v_j) \wedge v_j \in NN(v_i)$, possibly resulting in nodes having degree less than $k$.

- The $\varepsilon$-weighted graph is where only the subset of edges with weight greater than a threshold $\varepsilon$ from the fully connected graph are included.

The methods we describe do not make many assumptions about the nature of the graph or the edge weight distribution, beyond the fact that the weights on edges are assumed to be non-negative and $w_{ij} = 0$ if $(i, j) \notin E$.

**Types of Labels.** Users of social networks often reveal only partial information about themselves. For instance, only a subset of users reveal their age or location on Facebook. Therefore, the graph abstracted from user-generated data has labels on only a subset of nodes. The labels on nodes can be of different types:

- binary: only two possible values are allowed (gender is often restricted to male or female); equivalently, a label may only appear in the positive form ("smoker") and its absence is assumed to indicate the negative form ("non-smoker").
- numeric: the label takes a numeric value (age, number of views), Only values in some range may be allowed (age restricted to 0-120), or the range may be "bucketized" (age is 0-17; 18-35; 36-50, 50+).
- categorical: the label may be restricted to a set of specified categories (such as for interests, occupation).
- free-text: users may enter arbitrary text to identify the labels that apply to the node (as in listing favorite bands, or tags that apply to a photograph).

Some datasets have many labels in many categories (such as age, gender, location, and interests in a Facebook dataset), while in some cases there may be only a single label recorded (e.g. the only profile information available about a user in Netflix is location). Some label types allow a single value (users may declare only a single age), while others allow many values (users may list many differing tags for a photograph). For node classification, our goal is typically to provide values of a single label type, even though we may have information about other label types to use as features: our goal may be to predict age of Facebook users, given (partial) information about other users' age, sex and location.

In some cases, there may be weights associated with labels. For instance, the video sharing service Hulu displays the number of times each tag was assigned to a video. When normalized, the label-weight vector at each node can be thought of as providing confidence scores, or in some cases as probabilities for different labels. This can arise even when the original data does not exhibit weights: when labels are imputed, these may be given lower confidence scores than labels that were present in the original data. Depending on the dataset, some labels may be known for all nodes (e.g., the number of videos posted by a user on YouTube), and can be used as additional features for inferring the missing values of other labels.

## 2.2 The Node Classification Problem

We can now formally define the node classification problem.

**Problem Statement.** We are given a graph $G(V, E, W)$ with a subset of nodes $V_l \subset V$ labeled, where $V$ is the set of $n$ nodes in the graph (possibly augmented with other features), and $V_u = V \setminus V_l$ is the set of unlabeled nodes. Here $W$ is the weight matrix, and $E$ is the set of edges. Let $\mathcal{Y}$ be the set of $m$ possible labels, and $Y_l = \{y_1, y_2, \ldots, y_l\}$ be the initial labels on nodes in the set $V_l$. The task is to infer labels $\tilde{Y}$ on all nodes $V$ of the graph.

**Preliminaries and Definitions.** Let $V_l$ be the set of $l$ initially labeled nodes and $V_u$ the set of $n - l$ unlabeled nodes such that $V = V_l \cup V_u$. We assume the nodes are ordered such that the first $l$ nodes are initially labeled and the remaining nodes are unlabeled so that $V = \{v_1, \ldots, v_l, v_{l+1}, \ldots, v_n\}$. An edge $(i, j) \in E$ between nodes $v_i$ and $v_j$ has weight $w_{ij}$. A transition matrix $T$ is computed by row normalizing the weight matrix $W$ as:

$$T = D^{-1}W$$

where $D$ is a diagonal matrix $D = \mathrm{diag}(d_i)$ and $d_i = \sum_j w_{ij}$. The *unnormalized graph Laplacian* of the graph is defined as: $L = D - W$, and the *normalized graph Laplacian* as: $\mathcal{L} = D^{-1/2} L D^{-1/2}$. If $W$ is symmetric, then both these Laplacians are positive semi-definite matrices.

Let $Y_l = \{y_1, y_2, \ldots, y_l\}$ be the initial labels from the label set $\mathcal{Y}$, on nodes in the set $V_l$. The label $y_i$ on node $v_i$ may be a binary label, a single label or a multi-label. For binary classification, we may distinguish the presence of a label as $y_i \in \{-1, 1\}$ if $v_i \in V_l$ and 0 otherwise to indicate the absence of a label. For a single label classification, $y_i$ can take values from $\mathcal{Y}$, the range of possible values of that label. Finally, for a multi-class classification, $y_i$ denotes a probability distribution over $\mathcal{Y}$, where $\mathcal{Y}$ is the set of possible labels. For any label $c \in \mathcal{Y}$, $y_i[c]$ is the probability of labeling node $v_i$ with label $c$. Here, $Y_l$ is a matrix of size $l \times m$. We denote the initial label matrix of size $n \times m$ as $Y$, such that it has the first $l$ rows as $Y_l$ for labeled nodes and zeros in the next $n - l$ rows representing unlabeled nodes.

The output of the node classification problem is labels $\tilde{Y}$ on all nodes in $V$. A slightly different problem is to determine the labels on only the unlabeled nodes (i.e., $\tilde{Y}_u$ for nodes in $V_u$), and assume that the labels on the nodes in $V_l$ are fixed. Another variant is to learn a labeling function $f$ which is used to determine the label on the nodes of the graph.

## 3. Methods using Local Classifiers

In the following sections we describe the different approaches to solve the node classification problem and its variations. We start by describing a class of iterative methods that use local neighborhood information to generate features that are used to learn local classifiers.

These iterative methods are based on building feature vectors for nodes from the information known about them and their neighborhood (immediately adjacent or nearby nodes). These feature vectors are then used along with the known class values $Y_l$, to build an instance of a local classifier such as Naïve Bayes, Decision Trees etc. for inferring the labels on nodes in $V_u$.

### 3.1 Iterative Classification Method

This notion of iteratively building a classifier is quite natural and has appeared several times in the literature. Here, we follow the outline of Neville and Jensen [26].

**Input.** As with traditional classification problems, a set of attributes may be known for each node $v_i \in V$. These attributes form a part of the feature vector for that node and are known as *node features*. Consider the YouTube graph from Example 1.3: attributes such as the number of times a video is viewed, the time of upload, rating etc. are node features that are known for each video.

What makes classification of graph data different is the presence of links between objects (nodes) being classified. The information about the neighborhood of each object is captured by *link features*, such as the (multi)set of tags on videos. Typically, link features are presented to the classifier as aggregate statistics derived from the labels on nodes in

**Algorithm 1:** ICA($V, E, W, Y_l$)

Compute $\Phi^1$ from $V$, $E$, $W$, $Y_l$
Train classifier using $\Phi_l$
**for** $t \leftarrow 1$ **to** $\tau$ **do**
  Apply classifier to $\Phi_u^t$ to compute $Y_u^t$
  Update $\Phi_u^t$
$\tilde{Y} \leftarrow Y^\tau$
**return** $\tilde{Y}$

the neighborhood. A popular choice for computing a link feature is the frequency with which a particular label is present in the neighborhood. For instance, for a video $v_i$ in the YouTube graph, the number of times the label *music* appears in the nodes adjacent to $v_i$ is a link feature. If the graph is directed, the link features may be computed separately for incoming and outgoing links. The features may also include graph properties, such as node degrees and connectivity information.

**Iterative Framework.** Let $\Phi$ denote the matrix of feature vectors for all nodes in $V$, where the $i$-th row of $\Phi$ represents the feature vector $\phi_i$ for node $v_i$. The feature vector $\phi_i$ may be composed of both the node and link features; these are not treated differently within the vector. Let $\Phi_l$ and $\Phi_u$ denote the feature vectors for labeled and unlabeled nodes respectively. Algorithm 1 presents the Iterative Classification Algorithm (ICA) framework for classifying nodes in a graph. An initial classifier is trained using $\Phi_l$ and the given node labels $Y_l$. In the first iteration, the trained classifier is applied to $\Phi_u$ to compute the new labeling $Y_u^1$. For any node $v_i$, some previously unlabeled nodes in the neighborhood of $v_i$ now have labels from $Y_u^1$. Since link features are computed using the labels in the neighborhood, after applying the classifier once, the values of these features can change. It therefore makes sense to iterate the ICA process. In the $t$th iteration, the procedure builds a new feature vector $\Phi^t$ based on $\Phi_l$ and $Y_u^{t-1}$, and then applies the classifier to produce new labels $Y_u^t$. Optionally, we may choose to retrain the classifier at each step, over the current set of labels and features.

If node features are not known, the inference is based only on link features. In such a case, if a node has no labeled node in its neighborhood, it is remains unlabeled in the first iteration. As the algorithm proceeds, more nodes are labeled. Thus, the total number of iterations $\tau$ should be sufficiently large to at least allow all nodes to receive labels. One possibility is to run the iteration until “stability” is achieved, that is, until no label changes in an iteration—but for arbitrary local classifiers there is no guarantee that stability will be reached. Instead, we may choose to iterate for fixed number of iterations that is considered large enough, or until some large fraction of node labels do not change in an iteration.

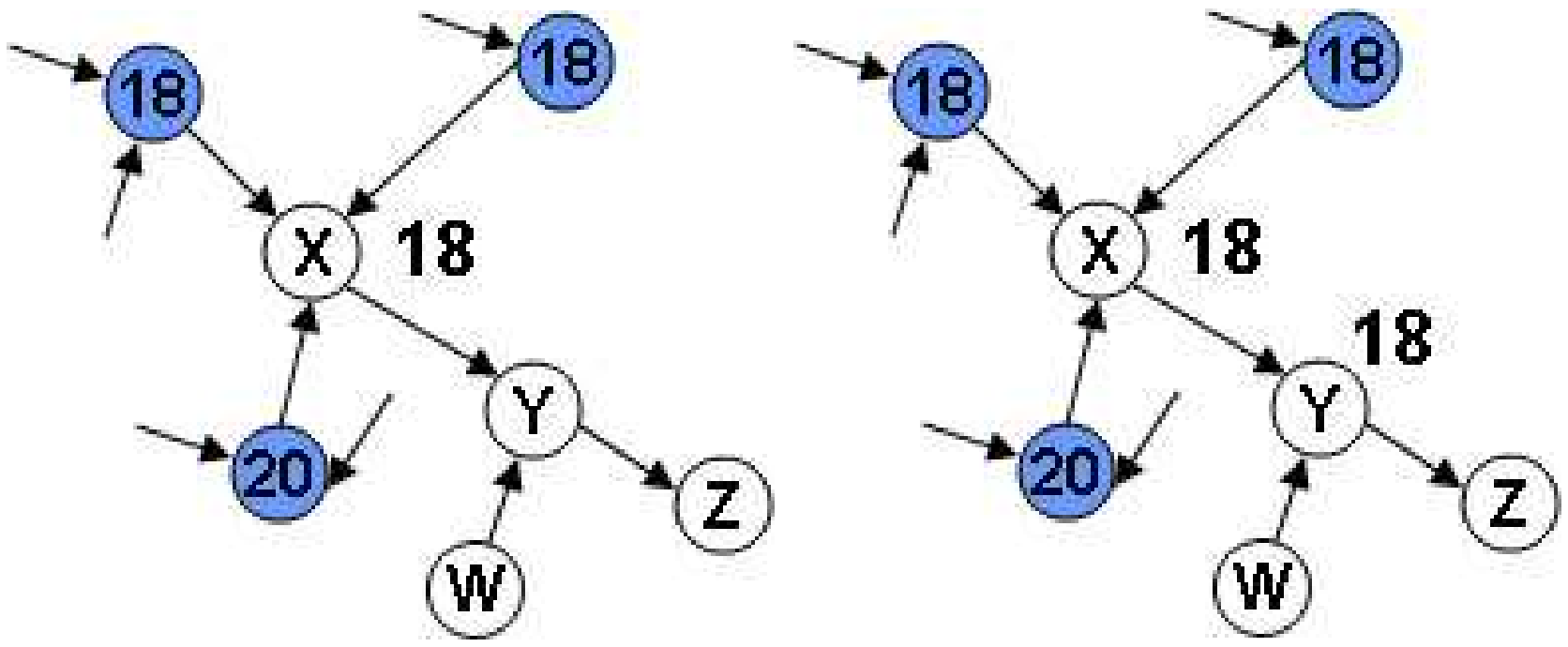

(a) Step 1 (b) Step 2

*Figure 1.1.* Two steps of a local iterative approach to node classification

Figure 1.1 shows two steps of local iteration on a simple graph. Here, shaded nodes are initially labeled. In this example, the first stage labels node X with the label '18'. Based on this new link feature, in the second iteration this label is propagated to node Y. Additional iterations will propagate the labeling further.

A consequence of basing the labeling solely on the labels of other nodes (a common characteristic of many of the methods we describe in this chapter) is that if the graph contains isolated components that do not have a single labeled node, then all nodes in that component will remain unlabeled, no matter how many iterations are applied.

**Instances of the Iterative Framework.** Neville *et al.* originally used a Naïve Bayes classifier to infer labels in their instantiation of the ICA framework [26]. A strategy they found useful was to sort the predicted class labels in descending order of their associated probability and retain only the top-$k$ labels, thus removing the less confident possibilities. Since then, ICA has been applied to graph data from different domains, with a variety of classifiers. For example, Lu and Getoor [20] applied logistic regression to classify linked documents.

An important special case is the method of Macskassy and Provost [21], who used a simpler classification method based on taking a weighted average of the class probabilities in the neighborhood (effectively "voting" on the label to assign). This classifier is based on a direct application of homophily (the premise that nodes link to other nodes with similar labels), and uses the immediate neighborhood of a node for classification. Bhagat *et al.* [5] proposed a method that considers the labeled nodes in the entire graph. This can be viewed as an instance of ICA using a nearest neighbor classifier to find a labeled node that is most similar to an unlabeled node being classified. It is based on co-citation regularity, the premise that nodes with similar neighborhoods have similar labels. These two simple methods (voting and nearest neighbor) are shown to be surprisingly effective on social network data, achieving quite high accuracy (70-90% on certain demographic labels) from relatively little local information [5].

One of the seminal works on classification of linked documents was by Chakrabarti *et al.* [8]. Their method used features from neighboring documents to aid the classification, which can be viewed as an instance of ICA on a graph formed by documents. Their experiments showed a significant improvement when using link features over just using the text at each node.

## 4. Random Walk based Methods

The next set of methods we discuss are based on propagating the labels by performing random walks on the graph. These are often thought of as semi-supervised learning or transductive learning methods and can be shown to be equivalent to learning a global labeling function over the graph with provable convergence guarantees. Unlike the iterative methods described so far that rely on computing link features to encode the information in the neighborhood, these methods more explicitly use the link structure for labeling nodes. However, we will see that there are strong connections between random walk and iterative methods.

The idea underlying the random walk methods is as follows: the probability of labeling a node $v_i \in V$ with label $c \in \mathcal{Y}$ is the total probability that a random walk starting at $v_i$ will end at a node labeled $c$. The various methods proposed in the literature differ in their definition of the random walk used for labeling. For this to provide a complete labeling, the graph $G$ is often assumed to be *label connected* [2]. That is, it is possible to reach a labeled node from any unlabeled node in finite number of steps.

The random walk is defined by a transition matrix $P$, so that the walk proceeds from node $v_i$ to node $v_j$ with probability $p_{ij}$, the $(i, j)$-th entry of $P$. For this to be well defined, we require $0 \leq p_{ij} \leq 1$ and $\sum_j p_{ij} = 1$. The matrix $P$ also encodes the *absorbing states* of the random walk. These are nodes where the state remains the same with probability 1, so there is zero probability of leaving the node, i.e., if a random walk reaches such a node, it ends there. Let $p_{ij}^{(t)}$ be the probability of reaching node $v_j$ after $t$ steps of a random walk that started at node $v_i$, and let $P^t$ denote the corresponding matrix at time $t$. For $t \to \infty$, the entry $p_{ij}$ of matrix $P^\infty$ represents the probability of the walk that starts at node $v_i$ is at node $v_j$ as the length of the walk tends to infinity. That is, if the *start distribution* is $e_i$, the vector with 1 at the $i$-th position and zeros elsewhere, a random walk on $G$ with transition matrix $P$ will converge to a stationary distribution which is the $i$-th row of the matrix $P^\infty$. We will often be able to show a closed-form expression for $P^\infty$ as a function of $P$.

**Labeling.** The walk is typically defined over nodes of the graph, and this is used to define a labeling. The probability of label $c \in \mathcal{Y}$ being assigned to node $v_i$ is computed as the total probability of reaching nodes labeled $c$ on convergence, starting at $v_i$. More precisely,

$$\tilde{y}_i[c] = \sum_{j|v_j \in V_l} p_{ij}^\infty y_j[c] \tag{1.2}$$

where the input label $y_j$ at $v_j \in V_l$ is assumed to be a probability distribution over labels. If the graph is label connected, as $t \to \infty$ the probability of reaching a labeled node is 1, so it follows that the output labeling $\tilde{y}_i$ at node $v_i \in V$ is also a probability distribution with

$$\sum_{c \in \mathcal{Y}} \sum_{j|v_j \in V_l} p_{ij}^\infty y_j[c] = 1.$$

If the output is required to be a single label on each node, then the most probable label can be assigned to each node, i.e.

$$\tilde{y}_i = \arg\max_{c \in \mathcal{Y}} \sum_{j;y_j=c} p_{ij}^\infty.$$

Recall that $Y$ is the matrix that records the label distribution for each labeled node, and 0 for unlabeled nodes. Consequently, the matrix equation for node classification using random walks can be written as:

$$\tilde{Y} = P^\infty Y \tag{1.3}$$

When $Y$ represents a matrix of probability distributions over the label set, then the result can be scaled to ensure that the output is a probability distribution as: $\tilde{Y} = N^{-1} P^\infty Y$, where $N^{-1}$ is a diagonal normalization matrix, defined as $N_{ii} = \sum_{j=1}^{m} (P^\infty Y)_{ij}$. We next consider various methods based on random walks in detail. Given a description of a random walk, we aim to find a description of the stationary distribution $P^\infty$, from which the labeling follows using the above equations.

## 4.1 Label Propagation

The node classification method of Zhu *et al.* [38] was proposed in the context of semi-supervised learning, where a symmetric weight matrix $W$ is constructed using Equation (1.1). More generally, we consider it to take as input a graph $G(V, E, W)$, from which we derive the matrix $T = D^{-1}W$. Nodes $V_l$ have initial labels $Y_l$ from the label set $\mathcal{Y}$.

**Random Walk Formulation.** The random walk at node $v_i$ picks an (outgoing) edge with probability proportional to the edge weight, if $v_i$ is unlabeled; however, if $v_i$ is labeled, the walk always loops to $v_i$. Therefore the nodes in $V_l$ are absorbing states, i.e. they are treated as if they have no outgoing edges, and thus their labels do not change. Since we have ordered the nodes so that the labeled nodes are indexed before the unlabeled nodes, we can write the transition matrix $P$ as a block matrix,

$$P = \begin{pmatrix} P_{ll} & P_{lu} \\ P_{ul} & P_{uu} \end{pmatrix} = \begin{pmatrix} I & 0 \\ P_{ul} & P_{uu} \end{pmatrix} \tag{1.4}$$

where the matrix $P_{ll}$ corresponds to the block of probabilities corresponding to transitions from a labeled node to another labeled node, and so on. Recall that a random walk that reaches a labeled node ends there, thus $P_{ll}$ reduces to $I$ and $P_{lu}$ is a matrix of all zeros. In other words, the transition matrix $P$ can be defined as $P_i = (D^{-1}W)_i$ if $i \in V_u$, else $P_i = e_i$ if $i \in V_l$.

Now, computing the matrix $\lim_{t\to\infty} P^t$, we obtain

$$P^\infty = \begin{pmatrix} I & 0 \\ (I - P_{uu})^{-1}P_{ul} & P_{uu}^\infty \end{pmatrix}. \tag{1.5}$$

For a graph in which every connected component has a labeled node, each entry of matrix $P_{uu}$ is less than 1. So $P_{uu}^\infty = 0$, and $P_{ul}^\infty$ is a distribution over labeled nodes. By combining Equations (1.3) and (1.5), the labels on unlabeled nodes can be computed as

$$\tilde{Y}_u = (I - P_{uu})^{-1}P_{ul}Y_l \tag{1.6}$$

This provides a precise characterization of how the labels on nodes are computed, and can be applied directly when $P_{uu}$ is small enough

**Algorithm 2:** LP-Zhu$(Y, P)$

$Y^0 \leftarrow Y$
**repeat**
  $Y^t \leftarrow PY^{t-1}$
  $Y_l^t \leftarrow Y_l$
**until** *convergence to* $Y^\infty$
$\tilde{Y} \leftarrow Y^\infty$
**return** $\tilde{Y}$

to invert directly. The solution is valid only when $(I - P_{uu})^{-1}$ is non-singular, which is true for all label connected graphs.

Observe that if the labeled nodes were not defined to be absorbing states, a random walk over $G$ would converge to a stationary distribution that is independent of the starting point (and hence would not be meaningful for the purpose of labeling). Szummer and Jaakkola [31] considered the variation where the labeled nodes are not forced to be absorbing states. In their definition, they perform random walks for $t$ steps. Such methods depend critically on the parameter $t$: it is easy to see that the extreme case $t = 1$ gives exactly the local voting scheme of Macskassy and Provost executed for a single iteration, while we argued that allowing $t$ to grow too large, the process mixes, and is independent of the starting location. In experiments, Szummer and Jaakola found small constant values of $t$ to be effective, around $t = 8$ on a dataset with a few thousand examples.

**Iterative Formulation.** We now show that this random walk is equivalent to a simple iterative algorithm in the limit. Consider an iterative algorithm where each node is assigned a label distribution (or a null distribution) in each step. In step $t$, each unlabeled node takes the set of distributions of its neighbors from step $t-1$, and takes their mean as its label distribution for step $t$. The labels of the labeled nodes $V_l$ are not changed. This is essentially the iterative algorithm of Macskassy and Provost [21] described above. The initial label distributions are given by $Y$. We can observe that each iterative step has the effect of multiplying the previous distribution by $P$, the block matrix defined above. This is illustrated in Algorithm 2.

The $t$th iteration of the algorithm sets

$$Y_u^t = P_{ul} Y_l + P_{uu} Y_u^{t-1},$$

which can be rewritten as

$$Y_u^t = \sum_{i=1}^{t} P_{uu}^{i-1} P_{ul} Y_l + P_{uu}^t Y_u.$$

On convergence, we get

$$\tilde{Y}_u = \lim_{t \to \infty} Y^t = (I - P_{uu})^{-1} P_{ul} Y_l.$$

In other words, this iterative algorithm converges to the same labeling as the random walk just described.

Thus, node classification performed by the iterative Algorithm 2 is the same as solving the matrix equation (1.6). Other equivalences can be shown: for an input graph $G$ which has a symmetric weight matrix $W$ and binary labels $\mathcal{Y} = \{0, 1\}$ on nodes, Zhu *et al.* show that labeling using equation (1.6) is equivalent to computing $\tilde{Y}$ using the value of a minimum energy harmonic function $f : V \to \mathbb{R}$ [38].

**Rendezvous approach to Label Propagation.** Azran [2] showed a different analysis of the label propagation random walk that more strongly uses the assumption that the graph is label connected. Since $G$ is label connected, the probability of moving from one unlabeled node to another after an infinite number of steps is 0, i.e., $P_{uu}^{\infty}$ is a zero matrix. Therefore, the limiting matrix $P^{\infty}$ has the form:

$$P^{\infty} = \begin{pmatrix} I & 0 \\ (P^{\infty})_{ul} & 0 \end{pmatrix} \tag{1.7}$$

Let $P = S \Lambda S^{-1}$ and $P^{\infty} = S \Lambda^{\infty} S^{-1}$, where $S$ is the matrix of left eigenvectors of $P$ and $\Lambda$ is the diagonal matrix of eigenvalues. Azran showed that the structure of $P$ and $P^{\infty}$ is such that the $l$ leading eigenvalues of both are 1. All other eigenvalues of $P$ have magnitude less than 1, and so are negligible for $P^{\infty}$. Thus, to compute $P^{\infty}$, it is sufficient to compute the $l$ leading eigenvectors of $P$. Further, the transition matrix defined in Equation (1.7) can be computed as:

$$(P^{\infty})_{ij} = \frac{s_{ij}}{s_{jj}} \tag{1.8}$$

where $i, j$ are indices for unlabeled and labeled nodes respectively. That is, the equation holds for $l + 1 \le i \le n$ and $1 \le j \le l$. As before, the labels on unlabeled nodes can then be computed using Equation (1.2).

In many applications, the number of unlabeled nodes is typically much larger than the number of labeled nodes. Thus, while inverting the matrix $(I - P_{uu})$ from the label propagation method would be too expensive, computing the $l$ principal eigenvectors of $P$ may be cheaper.

**Algorithm 3:** LP-ZHOU$(Y, \alpha, T)$

$t \leftarrow 1$
$Y^0 \leftarrow Y$
**repeat**
  $Y^t \leftarrow \alpha T Y^{t-1} + (1-\alpha) Y^0$
**until** *convergence to* $Y^\infty$
**foreach** $i \in V$ **do**
  $c = \arg\max_{j \in \mathcal{Y}} y_i^\infty[j]$
  $\tilde{y}_i[c] = 1$
**return** $\tilde{Y}$

## 4.2 Graph Regularization

The graph regularization method introduced by Zhou *et al.* [36] differs from label propagation in a key way: the labels on nodes in $V_l$ are allowed to change during label propagation. The initial labels are represented by a binary $n \times m$ matrix $Y$ such that $y_i[c] = 1$ if node $v_i$ has label $c \in \mathcal{Y}$, and each node has at most one label.

We first describe the method in terms of a random walk starting from a node $v_i$. Now the random walk at every node proceeds to a neighbor with probability $\alpha$ (whether or not the node is unlabeled) but, with probability $1-\alpha$ the walk jumps back to $v_i$, the starting node. Here, $1-\alpha$ can be thought of a "reset probability". In matrix form, the $t$-step transition probability $Q^t$ can be written as

$$Q^t = \alpha T Q^{t-1} + (1-\alpha) I.$$

This random walk has been well studied in other contexts—in particular, it corresponds to the "personalized page rank", introduced by Jeh and Widom [14] It can be shown to converge, to the stationary distribution

$$Q^\infty = (1-\alpha)(I - \alpha T)^{-1}.$$

Further, the corresponding label distribution can be computed as

$$\tilde{Y} = Q^\infty Y = (1-\alpha)(I - \alpha T)^{-1} Y.$$

**Iterative Formulation.** As in the previous case, this can also be seen as implementing a simple iterative method: at each step, the label distribution of node $i$ is computed as an $\alpha$ fraction of the sum of label distributions of its neighbors from the previous step, plus a $1-\alpha$ fraction of its initial label distribution. This is illustrated in Algorithm 3. One can verify that this formulation leads to the same solution, i.e. the final label distribution is given by

$$\tilde{Y} = Y^{\infty} = (1-\alpha)(I-\alpha T)^{-1}Y \tag{1.9}$$

which can be scaled up appropriately (via a diagonal normalization matrix) so there is a probability distribution over labels.

**Regularization Framework.** Zhou *et al.* considered several variations of this method, based on replacing $P$ with related matrices. In particular, they suggest the symmetrically normalizing $W$ via the diagonal matrix of row sums $D$ as, $\mathcal{P} = D^{-1/2}WD^{-1/2}$. $\mathcal{P}$ can also be written in terms of the normalized Laplacian as $\mathcal{P} = I - \mathcal{L}$. This is no longer a stochastic matrix, since the rows do not yield probability distributions. Nevertheless, we can still apply this as a generalized random walk/iterative method by computing $Y^t = \alpha\mathcal{P}Q^{t-1}Y + (1-\alpha)Y$, which converges to $Q^{\infty} = (1-\alpha)(I-\alpha\mathcal{P})^{-1}Y$.

The choice of $\mathcal{P}$ can be seen as arising naturally from some requirements on the labeling. Two such requirements are: (1) the difference between initial and output labels on labeled nodes should be small; and (2) the difference in the labels of neighbors (especially those with large edge weight) should be small, i.e., neighbors should have similar labels.

The task of finding a labeling which satisfies these two conditions can be formulated as an optimization problem to find a function $\tilde{f}$ that minimizes the above two conditions. This process is known as "regularization". Formally, define

$$\tilde{f} = \arg\min_f \frac{\mu}{2}\|f - Y\|^2 + f^T\mathcal{L}f,$$

for a parameter $\mu > 0$. The first term measures the difference between the labeling given by the labeling function and the original labeling $Y$, to capture (1). The second term uses the Laplacian $\mathcal{L}$ (which is related to the gradient of the function) to measure the smoothness of the labeling function, to capture (2). The parameter $\mu$ then controls the tradeoff between these two terms. Thus, solving for $\tilde{f}$ is intended to satisfy both the requirements. Rewriting this minimization in terms of the Euclidean norm results in

$$\min_f \frac{1}{2}\sum_{i,j=1}^{n} w_{ij}\left\|\frac{f_i}{\sqrt{d_i}} - \frac{f_j}{\sqrt{d_j}}\right\|^2 + \frac{\mu}{2}\|f - Y\|^2 \tag{1.10}$$

Differentiating and equating to zero, we get

$$\tilde{f} - D^{-1/2}WD^{-1/2}\tilde{f} + \mu(\tilde{f} - Y) = 0$$
$$(1+\mu)\tilde{f} - \mathcal{P}\tilde{f} = \mu Y$$

Solving for $\tilde{f}$ and setting $\alpha = \frac{1}{1+\mu}$, we have

$$\tilde{f} = (1-\alpha)(I - \alpha\mathcal{P})^{-1}Y \tag{1.11}$$

Notice that if we defined the transition matrix $Q$ in terms $\mathcal{P}$ instead of $T$, or iterated over $\mathcal{P}$ instead of $T$, the solution on convergence would match Equation (1.9). In other words, we can argue that this solution is a natural consequence of the formalization of the two requirements (1) and (2). It is possible to motivate other node classification algorithms based on similar optimization criteria. The survey of Bengio *et al.* [4] has more details on this perspective.

## 4.3 Adsorption

The "adsorption" method, proposed by Baluja *et al.* [3] is also based on iteratively averaging the labels from neighbors, in common with the previous algorithms studied. However, this method incorporates some additional features and parameters, so it can be seen as a generalization of other methods.

Adsorption takes as input a directed graph $G$ with weight matrix $W$. The initial labels are represented as $Y = \{y_1, y_2, \ldots, y_n\}$ such that $y_i$ is the probability distribution over labels $\mathcal{Y}$ if node $v_i \in V_l$, and is zero if node $v_i \in V_u$. As with graph regularization, adsorption does not keep the labels on nodes in $V_l$ fixed, but instead lets them be set by the labeling process. In order to maintain and propagate the initial labeling, adsorption creates a *shadow* vertex $\tilde{v}_i$ for each labeled node $v_i \in V_l$ such that $\tilde{v}_i$ has a single incoming edge to $v_i$, and no outgoing edges. In other words, the shadow vertex is an absorbing state when we view the algorithm as a random walk. Then, the label distribution $y_i$ is moved from $v_i$ to the corresponding shadow vertex $\tilde{v}_i$, so initially $v_i$ is treated as unlabeled. The set of shadow vertices is $\tilde{V} = \{\tilde{v}_i | v_i \in V_l\}$.

The weight on the edge from a vertex to its shadow is a parameter that can be adjusted. That is, it can be set so that the random walk has a probability $1 - \alpha_i$ of transitioning from vertex $v_i$ to its shadow $\tilde{v}_i$ and terminating. This *injection probability* was set to be a constant such as $\frac{1}{4}$ for all labeled nodes (and 1 for all unlabeled nodes) in the initial experimental study [3].

**Random Walk Formulation.** Based on this augmented set of nodes and labels, the Adsorption method defines additional matrices.

First, $A$ captures the injection probabilities from each node $v_i$: $A$ is the $n \times n$ diagonal matrix $A = \mathrm{diag}(\alpha_1, \alpha_2, \ldots, \alpha_l, 1, \ldots, 1)$ where $1 - \alpha_i$ is the (injection) probability that a random walk currently at $v_i$ transitions to the shadow vertex $\tilde{v}_i$ and terminates. Hence $\alpha_i$ is the probability that the walk continues to a different neighbor vertex.

A transition matrix $T$ encodes the probability that the walk transitions from $v_i$ to each of its non-shadow neighbors, so $T = D^{-1}W$ as before. Consequently the transitions among the non-shadow vertices are given by $(AT)$, while the transitions to shadow vertices are given by the first $l$ columns of $(I - A)$, which we denote as $(I - A)_{(l)}$. Putting these pieces together, we obtain an overall transition matrix $R$ over the set of $l + n$ nodes $\tilde{V} \cup V$ as:

$$R = \begin{pmatrix} I & 0 \\ (I - A)_{(l)} & AT \end{pmatrix} \tag{1.12}$$

The first $l$ columns of $R$ represent the transition probabilities to $\tilde{V}$, and the next $n$ columns give the transition probabilities to (within) $V$. The $(l + i)$-th row of $R$ gives the probability of starting at node $v_i$ for $1 \leq i \leq n$, and moving to the other nodes of the graph in one step. We can compute $R^t$ to give the $t$-step transition probabilities. Assuming that the graph is label connected, and since $0 \leq \alpha_i \leq 1$ and $T$ is a row stochastic matrix, as $t \to \infty$ we get

$$R^{\infty} = \begin{pmatrix} I & 0 \\ (I - AT)^{-1}(I - A)_{(l)} & 0 \end{pmatrix} \tag{1.13}$$

Let $Y_s$ be the matrix of labels on shadow vertices, i.e, the labels that were originally associated with nodes in $V_l$, then the matrix of initial labels is defined as: $\bar{Y}^0 = \begin{pmatrix} Y_s \\ 0 \end{pmatrix}$. Then the labeling at step $t$ is $\bar{Y}^t = R^{t-1}\bar{Y}^0$, and as $t \to \infty$,

$$\bar{Y}^{\infty} = R^{\infty}\bar{Y}^0 = \begin{pmatrix} Y_s \\ (I - AT)^{-1}(I - A)_{(l)}Y_s \end{pmatrix} \tag{1.14}$$

It can be verified that $\bar{Y}^{\infty}$ from Equation (1.14) is an eigenvector of $R$ with eigenvalue 1. The output as defined by Equation (1.14) is also a linear combination of the initial labels. Since $R$ and its powers are row stochastic matrices, and the initial labels are probability distributions, the output at each node is guaranteed to be probability distribution over labels. The resulting labeling can be rewritten in terms of the original graph (without shadow vertices) as:

$$\tilde{Y} = (I - AT)^{-1}(I - A)_{(l)}Y_l \tag{1.15}$$

**Iterative Formulation.** We now describe a local averaging method that is equivalent to the random walk method described above. Consider the graph $G$ and weight matrix $W$ as above, but with the directionality of all edges reversed. At each iteration, the algorithm computes the label distribution at node $v_i$ as a weighted sum of the labels in the neighborhood, and the node's initial labeling, the weight being given by $\alpha_i$. Formally, at the $t$-th iteration, for a node $v_i \in V$, the label distribution is computed as

$$y_i^t = \alpha_i \sum_j p_{ji} y_j^{t-1} + (1 - \alpha_i) y_i^0 \tag{1.16}$$

Rewriting Equation (1.16) as a matrix equation, with $Y^0 = \begin{pmatrix} Y_l \\ 0 \end{pmatrix}$, and $A = \mathrm{diag}(\alpha_1, \alpha_2, \ldots, \alpha_l, 1, \ldots, 1)$ as before,

$$\begin{aligned} Y^t &= ATY^{t-1} + (I - A)Y^0 \\ &= (AT)^{t-1}Y^0 + \sum_{i=0}^{t-1} (AT)^i (I - A) Y^0 \end{aligned}$$

We know that $0 \leq \alpha_i \leq 1$, and $T$ is a stochastic matrix, thus as $t \to \infty$, we reach

$$\tilde{Y} = Y^\infty = (I - AT)^{-1}(I - A)Y^0 \tag{1.17}$$

**Connection to other methods.** Observe that Equations (1.17) and (1.15) can be made to agree, since $Y^0$ has first $l$ rows non-zero and remaining $u$ rows as zeros. In particular, (1.9) can be obtained from (1.17) by setting $A_{ii} = \alpha$ for all $i$. In other words, the graph regularization method can be viewed as a special case of adsorption. This can be seen by comparing the description of the iterative formulations of both, and observing that both rely on averaging neighborhoods with the original label of a node. However, the definition of adsorption prefers to set $\alpha_i$ to be 1 for unlabeled nodes, to ensure that the final labeling is directly a probability distribution over labels without rescaling.

A second equivalence is achieved by setting $\alpha_i = 0$ for all (initially) labeled nodes. This has the effect of making them absorbing (any random walk which reaches them halts there with probability 1) and so we obtain the original label propagation algorithm again. This analysis shows that the adsorption method unifies the previous random walk methods.

# 5. Applying Node Classification to Large Social Networks

Our motivation for node classification comes from social networks. These networks can have millions of nodes and billions of edges, and the label set may consist of thousands of labels, e.g., the set of all tags on videos in YouTube. As we have indicated in our discussion, it be very computationally expensive to directly apply some of the methods described to datasets of this size. In particular, several methods are described in terms of finding the solution to a set of matrix equations. For example, in the random walk based method by Zhou *et al.*, using Equation (1.9) requires inverting a matrix of size $n \times n$, where $n$ is the number of nodes in the graph. In general, inverting this matrix takes time $O(n^3)$, although for sparse matrices methods from numerical analysis aim to compute the inverse (approximately) more quickly [12]. Similarly, the method by Azran *et al.* requires computing the $l$ leading eigenvectors of an $n \times n$ matrix, which can be done in time $O(n^2)$ for a sparse matrix. When $n$ is large, costs that are significantly superlinear in $n$ are not feasible, and we look for more scalable solutions.

## 5.1 Basic Approaches

When matrix inversion is not practical, there are several alternate approaches to finding the stationary distribution of the random walks we consider:

**Iteration.** When the number of labels $m \ll n$, it can be more efficient to work with the iterative form of the random walk, applied to $Y$. That is, rather than compute the $(n \times n)$ stationary distribution via $P^\infty$, instead compute the $(n \times m)$ stationary distribution of label probabilities $Y^\infty$. This in turn is computed by iterating to compute $Y^1, Y^2, \ldots$. In the limit, this converges to $Y^\infty$.

In practice, we may not run this procedure to convergence, but rather for a sufficiently large fixed number of iterations, or until the difference $\|Y_u^t - Y_u^{t-1}\|$ is sufficiently small. Such power iteration methods are known to converge exponentially quickly (i.e. the difference $\|Y_u^t - Y_u^\infty\|$ decreases by a constant factor each iteration) [17]. Hence only a constant number of steps is needed to reach a close enough approximation—of the order of tens to a hundred iterations.

**Random Walk Simulation.** An alternative approach is to directly simulate $r$ random walks that begin at $v_i$ for some number of steps, and use the distribution of the end point of these random walks as a surrogate for the stationary distribution. In some situations, such as where the graph is too large to represent as a matrix and compute matrix multiplications with, it may be more efficient to instead simulate random walks, which just require to access the adjacency lists of visited nodes, to pick the next node. This has been applied when the data is too large to store in memory [28].

## 5.2 Second-order Methods

When the classification method can be formalized as a matrix iteration, as in the random walk methods, then there have been many approaches suggested to reducing the number of iterations needed before the process converges. The main idea behind these *second-order methods* is that the update performed at each iteration is adjusted by the update performed at the previous iteration. These methods have been shown to converge more rapidly than simple iterations (referred to as *first-order methods*) for applications such as load balancing [24] and multi-commodity flow [25]. For node classification, a first order iteration of the form $Y^{t+1} = PY^t$ can be reformulated as $Y^{t+1} = \beta PY^t + (1-\beta)Y^{t-1}$ to yield a second-order method. Here, $\beta$ is a parameter weighting the current update. Second order methods have been shown to converge faster for $1 \leq \beta \leq 2$.

## 5.3 Implementation within Map-Reduce

The Map-Reduce framework [9] is a popular programming model that facilitates distributing computation over a cluster of machines for data-intensive tasks. Applications in this framework are implemented via two operations (1) Map: input represented as key/value pairs is processed to generate intermediate key/value pairs, and (2) Reduce: all intermediate pairs associated with the same key are collected and aggregated. The system takes care of allocating map and reduce tasks to different machines, which can operate in parallel.

This framework is particularly powerful for cases when the data being processed is so large that it does not fit on one machine, but must be stored in a distributed file system, as may be the case for social network data. We observe that all of the methods discussed so far fit well into the map-reduce model: both local iterative methods and random walk methods can be implemented so that each node collects information about its neighbors (map), and applies some process to compute its new label distribution (reduce).

We illustrate the power of Map-Reduce to distribute computations over large social networks with an example implementation of iterative

**Algorithm 4:** MAP(Key $v_i$, Value $y_i^{t-1}$)

**Data**: $P$
**foreach** $v_j \in V|(i,j) \in E$ **do**
  `Emit(`$v_j, (y_i^{t-1}, p_{ij})$`)`

**Algorithm 5:** REDUCE(Key $v_j$, ValueIterator *labelWt* )

$vec_{1\times m} \leftarrow 0$
**foreach** $(label, wt) \in labelWt$ **do**
  $vec[label]$ += $wt$
$y_j^t \leftarrow \arg\max(vec)$
`Emit(`$v_j, y_j^t$`)`

graph labeling methods. Consider a simple instance of the ICA method, where the classifier is based on weighted voting on labels in the neighborhood, as described in [5]. More specifically, at each iteration of the method, the label assigned to a node is the weighted vote of labels on its neighbors. This can be thought of as an iterative message passing scheme, where each node passes its current label (weighted by the edge weight) as a message to each neighbor. Then, each node collects the messages received from its neighbors and combines then to compute its label, in this case using a voting function.

This process fits neatly into the Map-Reduce setting. Algorithms 4 and 5 describe the Map and Reduce operations performed at each iteration of node classification. As described in Section 2, $P$ is the normalized weight matrix and $Y$ is a vector of initial labels, so that $y_i$ is a single label initially assigned to node $v_i$, and $y_i^t$ is the label at the $t$-th iteration. $vec$ is a temporary vector to aggregate the weight received for each label. The Map function implements the message passing, where each node $v_i$ sends a message $(y_i^{t-1}, p_{ij})$ at iteration $t$. The Reduce function receives the messages as $labelWt$ and aggregates them to infer the label $y_j^t$ at a node $v_j$ at iteration $t$.

Likewise, many other iterative algorithms for node classification can be implemented in Map-Reduce. One round of Map-Reduce computes each iteration (equivalently, computes the product of a matrix and a vector). Thus only a moderate number of rounds are required to converge on the solution.

# 6. Related approaches

In this section, we survey some of the other approaches to the node classification problem.

## 6.1 Inference using Graphical Models

The area of Statistical Relational Learning (SRL) has emerged over the last decade. SRL is generally concerned with creating models of data which fully describes the correlations between the different objects that are described by the data. It therefore encompasses node classification as a central problem. Another example of a problem in SRL is edge prediction: creating a model which can be used to predict which new edges are likely to be formed in a graph.

Among many approaches proposed within the SRL framework, two that have been directly applied to the node classification problem are Probabilistic Relational Models (PRMs) [10, 33] and Relational Markov Networks (RMNs) [32]. Essentially, the two approaches learn a probabilistic graph model to represent the relational (graph) data. The model used is a Bayesian network (directed) for PRMs and a Markov network (undirected) for RMNs. The learnt models are then used to perform inference or labeling over the graphs.

Formally, the nodes $V$ in the given graph $G$ are represented by random variables $X$ which can take values from the set $\mathcal{Y}$. Let $X_l$ denote the observed random variables associated with labeled nodes and $X_u$ be the unobserved variables for the unlabeled nodes. The task is to determine the joint probability distribution $P(X_u|X_l)$, i.e., the probability distribution over the labels for each unobserved variable (unlabeled node), given the observed variables (labeled nodes), and use it for inference.

In PRMs, each random variable $x_i$ representing node $v_i$ is associated with a conditional probability distribution $\mathbb{P}(x_i|\text{parents}(x_i))$, where $\text{parents}(x_i)$ is the set of label assignments on nodes that have an outgoing edge to node $v_i$. In the case of RMNs, a pairwise Markov Random Field (MRF) is defined over the graph that is parametrized by a set of arbitrary non-negative functions known as clique potentials.

These relational models are represented by a joint probability distribution of label assignments over nodes of the graph. This stands in contrast to methods such as random walks for graph labeling, which do not make these models for the label of a node explicit, but rather which use an implicit model to compute the labeling. To use a relational model for node classification, we must compute the marginal probability distribution for each node. However, this is not a simple task, since there are correlations (mutual dependencies) between the distributions on the nodes, and so there is no compact closed form for these marginals.

A popular choice for approximate inference in relational models is loopy belief propagation (LBP). LBP is an iterative message passing algorithm, where a message sent from node $v_i$ to $v_j$ is the belief of $v_i$ on what the value of $v_j$ should be. Pragmatically, this seems similar in spirit to the iterative methods which pass messages in the form of distributions of label values. LBP does not guarantee convergence except for special cases such as when the graphical model is a tree. Nevertheless, it has been found to work well in practice [34].

## 6.2 Metric labeling

There has been much other work on the problem of labeling objects when there are some relationships known between them. A central example is the work by Kleinberg and Tardos [15] that describes the problem of *Metric Labeling*. Here, there is a collection of objects with pairwise relationships between them (so relationships can be modeled as a graph). Each object also has an initial label (for example, this could be a user's declared age in a setting where many users do not reveal their true demographics). There are therefore two forces at work in the labeling: to pick labels which are consistent with the assigned labels of neighboring objects (due to implicit assumption of homophily), and to pick labels which are consistent with the initial labels. Kleinberg and Tardos formalize this problem by defining two functions, the first defined over (initial label, assigned label) pairs, and the second defined over (assigned label, neighbor's assigned label) pairs.

This then naturally defines an optimization problem over graphs: given the initial labeling of nodes, the edges, and the two cost functions, choose a labeling with minimum cost summed over all nodes and edges. Using the language of combinatorial optimization, this becomes a well-defined optimization problem, since every assignment has an associated cost: therefore, there must exist some assignment(s) with minimum cost. Kleinberg and Tardos are able to give guaranteed *approximations* to this problem: they show an algorithm to find an assignment whose cost is more expensive than the optimal one by a factor that is at most logarithmic in the number of labels. The approach relies on solving a carefully defined linear program, and arguing that rounding the fractional solution gives a good approximation to the original problem.

Applying this approach to the node classification problem is possible, but requires some effort. First, one needs to choose appropriate metrics over labels, and to extend these to capture the case of missing labels (which should not be penalized excessively by the metric over neighbors). Second, for graphs of social networks, it is necessary to solve a linear program defined over all the edges and nodes in the graph, which may stretch the limits of modern solvers.

## 6.3 Spectral partitioning

A different approach studied by McSherry [23] is to use spectral methods (study of eigenvalues and eigenvectors) to recover a labeling. For the analysis, the graph is assumed to have been produced by a random process. Each of the $n$ nodes in $V$ has a (secret) initial label from the $m$ possibilities. It is assumed that edges are created by a random process: let $Q$ be an $m \times m$ matrix where $q_{ij}$ denotes the probability of including an edge between a node with label $i$ and node with label $j$ for $i, j < m$. In this model, the given graph $G$ is drawn from the distribution implied by the hidden labels and $Q$. Computing the maximum-likelihood labeling is NP-hard in this setting, so McSherry presents an algorithm based on random partitioning and projections to infer the true labeling on the nodes of the graph with high probability.

Sidiropoulos (in the survey of Aggarwal *et al.* [1] points out that in most cases, it is not realistic to consider connections between all pairs of objects in the generative model. Hence the distribution of graphs considered by McSherry does not correspond to typical social network graphs. To better model such graphs, they consider a slightly modified model where for a given graph $H$, each edge $e$ of $H$ is *removed* with probability $1 - q_{ij}$, where $i, j$ are labels on the endpoints of $e$ and $Q$ is the $m \times m$ matrix of probabilities as before. Sidiropoulos argues that for arbitrary graphs $H$, it is not possible to recover almost all labels with high probability, but that simple algorithms may still recover a constant fraction of labels.

## 6.4 Graph Clustering

A set of approaches proposed for node classification are based on partitioning the nodes into clusters and assigning the same label to the nodes in the cluster. Blum and Chawla [7] assume that the weights on edges of the graph denote similarity between the associated nodes. Thus, a high edge weight means the nodes are very similar. A binary classification problem is solved by finding a "mincut": a partition of the nodes to minimize the number of edges crossing the cut. The intuition is that placing highly connected nodes in the same class will separate the nodes labeled "positive" from the "negative" ones. Such problems can be solved in polynomial time using classical max-flow algorithms.

This approach is motivated by observing that on certain cases, finding the minimum cut produces the smallest classification error.

Similar ideas are used by Shi and Malik [30], who propose a graph partitioning method for image segmentation. An image is represented as a graph, where a node represented a pixel in the image and the weight on an edge between two nodes represented the similarity between the pixel features (brightness, color etc.). After normalizing the edge weights, it is shown that the solution is equivalent to computing the second smallest eigenvector of a suitable matrix defined using the graph laplacian.

A recently proposed method by Zhou *et al.* [37] considers partitioning a graph based on both structural and attribute similarity of nodes of the graph. To combine the contribution of both types of feature, Zhou *et al.* define an *attribute augmented graph*, where new nodes are introduced for each attribute value. An edge is created between an existing node $v_i$ and an attribute node if $v_i$ has the corresponding attribute. To compute similarity between nodes, a distance measure based on a random walk is computed over the augmented graph, as in Section 4. However, rather than assign labels directly from the distribution of nodes reached, the method clusters nodes based on the random walk distance of pairs of nodes, via a method such as k-Medoids. Thus each cluster is assumed to consist of nodes that have similar labels: empirically, this was used to cluster (label) blogs based on political leaning, and researchers based on their research topic.

## 7. Variations on Node Classification

In this section, we identify a few generalizations of the graph labeling problem and methods which have been proposed for them.

### 7.1 Dissimilarity in Labels

Goldberg *et al.* [11] make the observation that nodes may link to each other even if they do not have similar labels. Consider the graph in Example 1.2, where an edge represents two videos that are often co-viewed. These videos may represent the two sides of a polarized debate, so the co-viewing relation could indicate that the videos are *dissimilar*, not similar. This can be formalized by allowing two types of edge, indicating either affinity or disagreement. Goldberg *et al.*assume that the type of each edge is known. For the case of binary labels, the goal is to find a function $f$ which maps nodes to either the +1 class or the -1 class. This is formalized as trying to minimize the following cost

function:

$$\sum_{i,j} w_{ij}(f(v_i) - s_{ij}f(v_j))^2$$

where $s_{ij}$ is 1 if edge $(i, j)$ is a similarity edge and -1 otherwise and $w_{ij}$ is the edge weight as before. This minimization requires solving a quadratic program to find $f$ to label the graph. It can be extended to the multi-class setting by introducing a labeling function $f_k$ for each class. Such programs can be solved for a moderate number of examples (hundreds to thousands) but may not scale to huge graphs.

## 7.2 Edge Labeling

Thus far in our discussion of node classification, we have assumed the weight matrix $W$ is known, or can be computed directly from node attribute similarity. But in fact the problem of computing edge weights can be abstracted as a problem in itself, to infer labels on edges of a graph. The simplest case is the binary classification problem to label the edges as positive or negative. For instance, in a network of blogs, a user may connect to users with whom they either (broadly) agree or disagree. Leskovec *et al.* [19] study the problem of classifying edges as positive and negative through the lens of two theories from social science literature: Balance and Status. The balance theory is based on notions such as "the friend of my friend is my friend", "the enemy of my friend is my enemy" etc. to balance the signs on edges within a triad. The status theory asserts that a positive (negative) edge from $v_i$ to $v_j$ indicates that the $v_i$ believes that $v_j$ has a higher (lower) status that $v_i$. Given two edges connecting three users, both models can predict the sign of the third edge, but disagree on some cases. Analyzing real data shows that balance tends to model the case of undirected edges, while status better captures the directed case.

Goyal *et al.* [13] studied a problem of edge labeling with applications such as viral marketing, where it is useful to know the influence that a user has on his neighbors. The problem is formulated as that of inferring a weight $0 \leq w_{ij} \leq 1$ on each edge $(i, j)$, termed as an influence probability. Specifically, influence is measured in terms of actions performed by neighbors of a user after the user has performed those actions. The authors present static and time-dependent models for learning edge weights.

Krushevskaja and Muthukrishnan [16] formulate a more general edge labeling problem, for arbitrary sets of possible edge labels. Formally, given a graph $G(V, E)$ with nodes $V$ and edges $E$, a subset of the edges $E_l \subset E$ are labeled. A label $y_i$ on edge $e_i \in E_l$ is a probability distri-bution over the label set $\mathcal{Y}$. The goal is to label all edges in the graph. As an example, consider a social network graph, with users represented by nodes and interactions between users represented by edges in the graph. The set of labels consists of types of interactions such as email, public post, tag in a photo, and video message. A subset of edges are labeled with a probability distribution over the label set. The probability associated with a label, say *email*, at any edge is the likelihood of the corresponding pair of users interacting by email. Krushevskaja and Muthukrishnan study two algorithms for the edge labeling problem, one in which edge labeling is reduced to node labeling on a line graph, and the other is a direct random walk based approach to edge labeling.

## 7.3 Label Summarization

In applications involving user generated data, such as tagging videos in YouTube (Example 1.3), a classification algorithm might choose a non-zero probability for a large number of labels on a given node. Simply picking the most likely few labels is not optimal: the chosen tags may be redundant, picking synonyms and omitting ones which better capture the content. A secondary concern is that computing the final label distribution is more costly if the algorithm has to store a complete label distribution for each node at each intermediate step. Given this motivation, Bhagat *et al.* [6] formulate a space-constrained variant of the graph labeling problem, where each node has space to store at most $k$ labels and associated weights. That is, given a partially labeled graph, the goal is to output a label summary of size $k$ at each node.

The simplest approach to solving the space-constrained labeling problem is to perform one of the node classification methods discussed earlier, and to prune each computed distribution to have only $k$ labels. A more sophisticated approach is to summarize label distributions using the semantic relationship among labels, modeled as a hierarchy over the label set. Now the goal is to choose $k$ appropriate labels from the hierarchy to best represent the computed distribution at each step. Such algorithms can run in small space, requiring space proportional to the number of labels in the neighborhood at any node.

# 8. Concluding Remarks

The problem of node classification has been defined and addressed in many works over the last 15 years, prompted by the growth of large networks such as the web, social networks, and other social media. In this chapter, we have surveyed the main lines of work, based on iterative methods and random walks, as well as several variations. When viewed from the right perspective, there is a surprising commonality between many methods: the methods discussed in Section 4, and several of the methods in Section 3 can all be seen as generating the labeling from the occupancy probabilities of a random walk over the graph. The universality of this concept, which may not have been obvious from the original papers, can be motivated from both linear algebraic and optimization considerations.

## 8.1 Future Directions and Challenges

Having surveyed so many approaches to this problem, we step back to ask the question, "Given a partially labeled graph, do we know how to classify the other nodes?". The short answer to this question is "no". Partly this is because the problem is underspecified: unless we make strong assumptions about the process that generates a true but hidden labeling, we are unable to prove any theorems which quantify the quality of any inferred labeling. So, as with other problems in the machine learning world, we should seek to evaluate solutions, by withholding some known labels and comparing the imputed labels on these nodes. Here, we enounter two limitations in the current literature. Firstly, some proposed methods do not scale up to the large graphs that result from the motivating social network setting, and have only been tested on graphs of a few thousand nodes or not at all. Secondly, for those techniques which do scale, there has been little attempt to compare multiple methods on the same baseline. A step in the right direction is the work of Macskassy and Provost [22], which compares multiple methods on relatively small data sets (hundreds to thousands of nodes).

Given these shortcomings, we identify the following challenges for the area:

- The random walk and iterative approaches have the advantage of being easy to implement and distribute via Map-Reduce. A next step is to provide baseline implementations of several of these methods, and to evaluate them across a common test set of large graphs derived from social networks to compare the behavior.

- Other methods suggested based on classical graph algorithms, combinatorial optimization, and spectral properties have not yet been tested on the large graphs which arise in from social networks. A natural next step is to understand how well approaches such as metric labeling, spectral partitioning and inference perform at this task, in terms of both accuracy and scalability. In particular, can they be implemented within the Map-Reduce framework?

- The complex models which arise from relational learning are typically solved by approximate message passing algorithms such as loopy belief propagation. Are there special cases of these models for which the LBP solution coincides with a known iterative algorithm? This would show a novel connection between these approaches.
- It is open to see if new algorithms can be developed which combine aspects from multiple different labeling methods to achieve a labeling which is better than that from any individual method (i.e. hybrid algorithms).
- Many methods are motivated based on hypotheses about what links are present: homophily, co-citation regularity, balance, status etc. To what extent can these hypotheses be tested within large scale data?

## 8.2 Further Reading

Several other sources have discussed different aspects of the graph labeling problem and its generalizations. As mentioned in Section 4.2, the survey by Bengio *et al.* [4] relates the semi-supervised learning methods in the context of optimizing a quadratic cost function.

The survey of Sen *et al.* [29] compares various node classification methods including ICA and relational learning method RMN that uses LBP. They empirically compare the methods on document classification tasks and show that the simple ICA method is the most efficient. The authors note that although in some experiments LBP had better labeling accuracy than ICA, but learning the parameters for the method is not trivial. Macskassy and Provost [22] survey different approaches, and give empirical comparisons on some web-based data sets. The tutorial of Neville and Provost [27] presents the machine learning perspective, and has slides and a reading list available.

## Acknowledgments

The work of the first and third authors is supported by NSF grant 0916782.